\documentstyle[twocolumn,prl,aps,epsfig]{revtex}

        \newcommand{\be}{\begin{equation}}
        \newcommand{\ee}{\end{equation}}
        \newcommand{\bea}{\begin{eqnarray}}
        \newcommand{\eea}{\end{eqnarray}}

\begin{document}                                                
%\draft
\title{Influence of the $\boldmath{U(1)_{A}}$ Anomaly on the QCD Phase Transition}
\author{Jonathan T. Lenaghan\thanks{Current address: The Niels Bohr Institute, Blegdamsvej 17, 
DK-2100 Copenhagen {\O}, Denmark}}
\bigskip
\address{
Department of Physics, Yale University,
New Haven, CT 06520-8124, USA\\
}
\date{\today}
\maketitle
\begin{abstract} 
The $SU(3)_{r} \times SU(3)_{\ell}$ linear sigma model is used 
to study the chiral symmetry restoring phase transition 
of QCD at nonzero temperature. The line of second order phase 
transitions separating the first order and smooth crossover 
regions is located in the plane of the strange and 
nonstrange quark masses.  
It is found that if the $U(1)_{A}$ symmetry 
is explicitly broken by the $U(1)_{A}$ anomaly 
then there is a smooth crossover to the chirally symmetric 
phase for physical values of the quark masses. 
However, if the $U(1)_{A}$ anomaly is absent,
the region of first order phase 
transitions is significantly enlarged and it is found that 
there is a phase 
transition for physical values of the quark masses 
provided that the $\sigma$ meson mass is 
at least 600 MeV.  
In both cases, the region of first order 
phase transitions in the quark mass plane 
is enlarged as the mass of the $\sigma$ meson is increased. 
\end{abstract} 
\pacs{}
\begin{narrowtext}

The ultimate goal of relativistic heavy ion experiments is 
to probe the phase diagram of Quantum Chromodynamics (QCD).
General theoretical considerations indicate that at 
sufficiently high temperatures there 
should be a transition from ordinary hadronic matter to 
a chirally symmetric plasma of quarks and gluons \cite{PisarskiWilczek}.
The order parameter for this phase transition is the 
quark-antiquark condensate.
Results from lattice gauge theory predict the temperature 
of this transition to be about 150 MeV \cite{generallattice}.  
The order of the phase transition, however, 
seems to depend very much on the number of quark flavors and 
their masses \cite{Columbia}.  

Classically, the matter part of the 
QCD Lagrangian with $N_f$ flavors is invariant under
the symmetry group
$SU(N_f)_{r} \times SU(N_f)_{\ell} \times U(1)_{A}$.
The axial $U(1)_{A}$ symmetry is broken to $Z(N_f)_A$ 
by a nonvanishing topological susceptibility \cite{thooft} and the 
$SU(N_f)_{r} \times SU(N_f)_{\ell}$ symmetry is 
spontaneously broken to the diagonal group of vector transformations, 
$SU(N_f)_{r+\ell}=SU(N_f)_{V}$, by a nonvanishing 
expectation value for the quark-antiquark condensate.  The 
$SU(N_f)_{r} \times SU(N_f)_{\ell} \times U(1)_{A}$ group is 
also explicitly broken by the effects of nonzero 
quark masses.  
It was shown by 
Pisarski and Wilczek that for three or more massless 
flavors, the phase transition for the 
restoration of the $SU(N_f)_{r} \times SU(N_f)_{\ell}$ symmetry 
is first order, while for two 
massless flavors the phase transition is 
second order\cite{PisarskiWilczek}.  

The $U(1)_A$ symmetry may also be 
restored, if only partially, since instanton effects
are Debye screened at high temperatures \cite{shuryak1,robyaffe}.
There are now two possibilities: either the $U(1)_A$ symmetry is 
restored at a temperature much greater than the 
$SU(N_f)_{r} \times SU(N_f)_{\ell}$ symmetry or 
the two symmetries are restored at (approximately) 
the same temperature \cite{shuryak}.  
Recent lattice gauge theory 
computations have demonstrated a rapid decrease in 
the topological susceptibility at $T_c$ \cite{digiacomo} and
random matrix models also indicate that the two 
symmetries are restored simulaneously \cite{zahed}.
Perhaps more dramatically, 
it was also shown that the topological susceptibility 
vanishes at $T_c$ in the large-$N_c$ limit \cite{dimarob}.
On the other hand, the fate of the $U(1)_A$ anomaly in nature 
is not completely clear since
instanton liquid model calculations indicate that 
the topological susceptibility is essentially unchanged at 
$T_c$ \cite{schafer}.
Additionally, other lattice computations which measure the 
chiral susceptibility find that the $U(1)_A$ symmetry restoration is 
at or below the 15\% level \cite{blumwingate,christ}. 

\begin{figure}
\hspace*{0.0cm}
\mbox{\epsfig{file=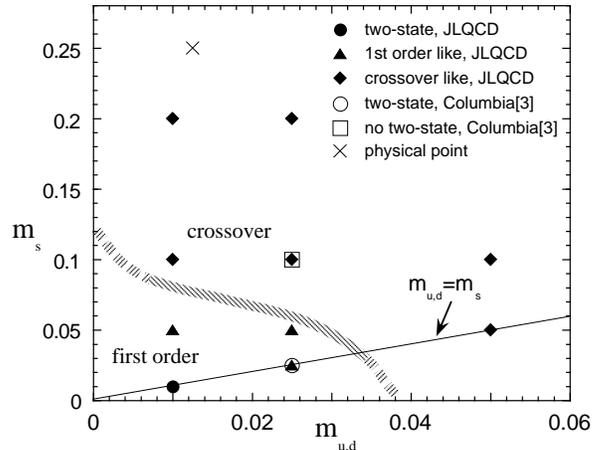,height=6cm,angle=0}}
\caption{The phase diagram on the ($m_{u,d}$,$m_{s}$) plane 
as obtained from lattice computations.  These results are a compilation 
of data from the JLQCD and the Columbia groups taken 
from Refs.\ [3] and [14].  The plot is from Ref.\ [14].
\label{fig:latticeplot}}
\end{figure}

Unlike the idealized massless quark limit, 
there are no general 
theoretical arguments which require that a phase transition 
exists for massive quarks.  
Indeed, some lattice simulations indicate that 
for physical quark masses, no phase 
transition occurs \cite{Columbia,lattice2,lattice3}.  
The general consensus from lattice
computations is that in the plane of 
light quark masses (see Fig.\ \ref{fig:latticeplot})
there is a first order region bounded 
by a line of second order transitions.  Outside this region, 
there is no phase transition, but rather a crossover characterized
by a rapid but smooth and continuous decrease of the 
quark-antiquark condensate.  
Given the present difficulties with performing lattice computations
with realistic quark masses and a large number of sites,
it is useful to complement the present lattice results with effective 
models that capture some of the relevant dynamics of QCD.  
Some work for three flavors has been done 
in this direction \cite{Hilde,Metzger,RobGavin,Pirner}. 
In these works, the $SU(3)_{r} \times SU(3)_{\ell}$ linear 
sigma model was used to study the order 
of the chiral symmetry restoring phase transition 
as a function of the current quark masses with the ratio 
of the up-down to strange quark masses held fixed.
In Refs.\ \cite{Hilde,Metzger,Pirner}, a loop-expansion is used to 
compute the effective potential and 
in Ref.\ \cite{RobGavin} a mean-field analysis of this model was 
presented.  
The effects of the restoration of the $U(1)_{A}$ symmetry 
on the spectrum of hadronic observables in heavy ion collisions 
was addressed within the context of this model in 
Ref.\ \cite{jurgenu1a}.

In this paper, I present
results concerning the order of the chiral symmetry 
restoring phase transition as a function of the 
current quark masses using the $SU(3)_{r} \times SU(3)_{\ell}$ linear 
sigma model without fixing the ratio of the masses.
In addition, the effects of the $U(1)_A$ anomaly on the 
order of the QCD phase transition are investigated.  
Here, the Cornwall-Jackiw-Tomboulis (CJT) \cite{CJT} formalism is 
used to derive gap equations for the condensates 
and the tadpole-resummed scalar and pseudoscalar nonet meson masses
at nonzero temperature.  
The derivation of and the solutions to this set of equations 
in a variety of limits 
has been presented elsewhere \cite{lrsb}.  
The results agree qualitatively with earlier 
studies on a lattice \cite{Columbia,lattice2,lattice3}
and with other studies using the $SU(3)_{r} \times SU(3)_{\ell}$ 
linear sigma model \cite{Hilde,Metzger,RobGavin,Pirner}. 
In the presence of an explicit $U(1)_{A}$ symmetry 
breaking term, I find that for physical values of the strange and 
nonstrange current quark masses, there is no
phase transition but rather a smooth 
crossover.  For smaller values of the masses, 
the phase transition is first order with a line of second order 
transitions separating the first order and the crossover 
regions.  
In the absence of an explicit $U(1)_{A}$ symmetry 
breaking term, the region of phase transitions 
is greatly enlarged.  In particular, 
if the $\sigma$ meson mass is greater than 600 MeV, then the 
transition is driven to first order for physical 
values of the quark masses.  In both cases, the 
region of first order phase transitions 
is enlarged as the mass of the $\sigma$ meson is increased.

The most general renormalizable theory compatible with the flavor 
symmetries of QCD is the $SU(3)_{r} \times SU(3)_{\ell}$ 
linear sigma model.
While this model cannot account for the full dynamics of QCD, 
on the line of second order phase transitions the only 
relevant dynamics are determined by the symmetries of the 
theory.  So, in the vicinity of this line, the use of the 
$SU(3)_{r} \times SU(3)_{\ell}$ linear sigma model is 
appropriate. 
Its Lagrangian is given by
\bea \label{L}
{\cal L}(\Phi) &=& 
{\rm Tr}  \left( \partial_{\mu} \Phi^{\dagger} 
\partial^{\mu} \Phi 
-  m^2 \, \Phi^{\dagger} 
\Phi \right) - 
\lambda_{1} \left[ {\rm Tr}  \left( \Phi^{\dagger}
 \Phi  \right) \right]^{2}  \nonumber \\
&-& \lambda_{2} {\rm Tr}  \left( \Phi^{\dagger} 
 \Phi  \right)^{2} 
+ c \left[ {\rm Det} \left( \Phi \right) + 
{\rm Det}  \left( \Phi^{\dagger} \right) \right]  \nonumber \\ 
&+& {\rm Tr} \left[H  (\Phi + \Phi^{\dagger})\right] \,\, . 
\eea
Here, $\Phi$ is a  $U(3)$ matrix defined by
$\Phi = T_a \, (\sigma_a + i \, \pi_a)$.
The $T_{a} = \hat{\lambda}_{a}/2$ are 
the generators of $U(3)$ where $\hat{\lambda}_{a}$ are 
the Gell-Mann matrices with 
$\lambda_{0} = \sqrt{2/3} \, I$. 
The $T_{a}$ are normalized such that 
${\rm Tr} (T_{a} T_{b}) = \delta_{ab}/2$.

The parameters of the Lagrangian are the bare mass $m$, 
a background matrix field $H = h_{0} 
T_{0} + h_{8}T_{8}$, a cubic coupling $c$ and 
two quartic couplings, $\lambda_{1}$ and $\lambda_{2}$. 
The various patterns of symmetry breaking and the 
parameterizations of the coupling constants for this Lagrangian 
were studied in \cite{lrsb} and will only be briefly 
reviewed here.  
For $\underline{H}=0$, $c=0$ and $m^2 > 0$, the Lagrangian has a global 
$SU(3)_{r} \times SU(3)_{\ell} \times U(1)_{A}$ symmetry.   
The effects of the $U(1)_{A}$ symmetry breaking by 
a nonvanishing topological susceptibility (i.e. the presence of 
instantons in the QCD vacuum) are
included by setting $c \neq 0$ which reduces the symmetry to
$SU(3)_{r} \times SU(3)_{\ell}$.
For nonzero $H$, chiral symmetry is explicitly broken.

I assume that 
there are nonzero vacuum expectation values for the $\sigma_{0}$ and 
$\sigma_{8}$ fields which I denote by ${\bar \sigma}_{0}$ 
and ${\bar \sigma}_{8}$.
After shifting these fields by their expectation values
and following \cite{ChanHay},
the Lagrangian can be 
rewritten as:
\bea
{\cal L}  &=& \frac{1}{2} [ \partial_{\mu} \sigma_{a} 
        \partial^{\mu} \sigma_{a} + \partial_{\mu} \pi_{a}
        \partial^{\mu} \pi_{a} - (m_{S}^{2})_{ab} \sigma_{a}
        \sigma_{b}  \nonumber \\
        &-&(m_{P}^{2})_{ab} \pi_{a} \pi_{b} ]  
         + ({\cal G}_{abc}-\frac{4}{3} \, {\cal F}_{abcd} \,
        {\bar \sigma}_{d})  \, \sigma_{a} 
	\sigma_{b} \sigma_{c}  \nonumber \\
        &-&3  \, ({\cal G}_{abc} + 
        \frac{4}{3}  \, {\cal H}_{abcd} \, {\bar \sigma}_{d} )
          \, \pi_{a} \pi_{b} \sigma_{c} 
        - 2  \, {\cal H}_{abcd}  \, \sigma_{a} \sigma_{b}
        \pi_{c} \pi_{d} \nonumber \\
        &-&\frac{1}{3}  \, {\cal F}_{abcd}  \, (
        \sigma_{a} \sigma_{b} \sigma_{c} \sigma_{d} + 
        \pi_{a} \pi_{b} \pi_{c} \pi_{d} ) - h_{a} {\bar \sigma}_{a} \,\, , 
\eea
where 
\begin{mathletters} 
\bea
{\cal G}_{abc} &=& \frac{c}{6} [ d_{abc} 
        - \frac{3}{2} \left(\delta_{a0} d_{0bc} +
        \delta_{b0} d_{a0c} + \delta_{c0} d_{ab0} \right) \nonumber\\
        &+& \frac{9}{2} d_{000} \delta_{a0} 
        \delta_{b0} \delta_{c0} ] \,\, , \nonumber \\
{\cal F}_{abcd} &=& \frac{\lambda_{1}}{4}  \, 
        \left(\delta_{ab} \delta_{cd} + 
        \delta_{ad} \delta_{bc} + \delta_{ac} \delta_{bd} \right) \nonumber \\
        &+&\frac{\lambda_{2}}{8} \, \left(d_{abn} d_{ncd} + 
        d_{adn} d_{nbc} + d_{acn} d_{nbd} \right) \,\, ,\nonumber\\
{\cal H}_{abcd} &=& \frac{\lambda_{1}}{4}  \, \delta_{ab} \delta_{cd}  + 
        \frac{\lambda_{2}}{8} \, (d_{abn} d_{ncd} + 
        f_{acn} f_{nbd}  \nonumber \\ 
        &+&  f_{bcn} f_{nad} ) \,\, , \nonumber \\
(m_{S}^{2})_{ab} &=& m^{2} \,  \delta_{ab} - 6 \,  {\cal G}_{abc} \, 
         {\bar \sigma}_{c} 
        + 4  \, {\cal F}_{abcd} \,  {\bar \sigma}_{c} {\bar \sigma}_{d} \,\, ,\nonumber \\
(m_{P}^{2})_{ab} &=& m^{2} \,  \delta_{ab} + 6  \, {\cal G}_{abc} \, 
         {\bar \sigma}_{c} 
        + 4  \, {\cal H}_{abcd}  \, {\bar \sigma}_{c} {\bar \sigma}_{d} \,\,.\nonumber
\eea
\end{mathletters}
Here the summation runs over the index $n$ only and 
$d_{abc}$ and $f_{abc}$ are the symmetric and 
antisymmetric structure constants, respectively, of $U(3)$.

The $\sigma_{a}$ fields are members of the 
scalar ($J^{\pi} = 0^{+}$) nonet and the $\pi_{a}$ 
fields are members of the pseudoscalar ($J^{\pi} = 0^{-}$) nonet.
The $\pi_{1,2,3}$ are the pions, the $\pi_{4,5,6,7}$ are the 
kaons and the $\pi_{0}$ and the $\pi_{8}$ are admixtures of 
the $\eta$ and the $\eta'$ with mixing angle $\theta_{{\rm P}}$.
The situation with the scalar nonet is not as clear and still 
somewhat controversial \cite{nonetnews}. 
The $\sigma_{0}$ and the $\sigma_{8}$ are admixtures
of the $\sigma$ and the $f_{0}(1370)$ with mixing angle 
$\theta_{{\rm S}}$.  The 
$\sigma_{1,2,3}$ are identified with the $a_{0}(980)$ and the 
$\sigma_{4,5,6,7}$ with the $\kappa$ meson.  

The explicit symmetry breaking terms can be determined 
(see, for instance, Ref.\ \cite{lrsb}), to be
\begin{mathletters}
\bea
h_{0} &=& \frac{1}{\sqrt{6}} \, (m_{\pi}^{2} \, f_{\pi} + 
        2 \, m_{K}^{2} \, f_{K}) \label{eqn:break1}\\
h_{8} &=& \frac{2}{\sqrt{3}} \, (m_{\pi}^{2} \, f_{\pi} -
m_{K}^{2} \, f_{K}) \,\, .\label{eqn:break2}
\eea
\end{mathletters}

The gap equations (Schwinger--Dyson equations) derived from 
the CJT effective potential \cite{CJT}
in the tadpole-resummed approximation, or Hartree 
approximation, are found to be
\bea
({\cal S}_{ab}&(k)&)^{-1} = -k^2+m^{2}
	\,\delta_{ab} - 6 \, {\cal G}_{abc} 	 
	\,\bar{\sigma}_{c} + 4 {\cal F}_{abcd}  
        \bar{\sigma}_{c} \, \bar{\sigma}_{d}\nonumber \\
        &+& 4 {\cal F}_{abcd}  \int_{k}  {\cal S}_{cd}(k) 
        +4\, {\cal H}_{abcd} \,
        \int_{k}  {\cal P}_{cd}(k) \,\, , \nonumber\\
({\cal P}_{ab}&(k)&)^{-1} = -k^2+m^{2} \, 
        \delta_{ab} + 6 \, {\cal G}_{abc} 
        \,\bar{\sigma}_{c} + 
        4 {\cal H}_{abcd}  
        \bar{\sigma}_{c} \, \bar{\sigma}_{d}  \nonumber \\
        &+& 4 {\cal H}_{abcd}  \int_{k}  {\cal S}_{cd}(k) 
         +4\, {\cal F}_{abcd} \,
        \int_{k}  {\cal P}_{cd}(k) \label{eqn:massgap2} \nonumber \\
h_{a} &=& m^{2} \, \bar{\sigma}_{a}  \,- 3 \, {\cal G}_{abc}
         \bar{\sigma}_{b} \, \bar{\sigma}_{c}  
        - 3 \, {\cal G}_{abc}\int_{k}  {\cal S}_{cb}(k) \nonumber \\ 
        &+& 3  \, {\cal G}_{abc}\int_{k} {\cal P}_{cb}(k) +
	\frac{4}{3} \, {\cal F}_{abcd} \bar{\sigma}_{d}\, 
        \bar{\sigma}_{b} \, \bar{\sigma}_{c} \nonumber \\ 
	&+& 4\, {\cal F}_{abcd} \bar{\sigma}_{d} 
	\int_{k} {\cal S}_{cb}(k) \, 
        + 4\, {\cal H}_{bcad} \,
        \bar{\sigma}_{d} \,  
        \int_{k}  {\cal P}_{cb}(k) \,\, , \label{eqn:cond}
\eea
where in the last equation, $a=0,8$ and ${\cal S}_{ab}(k)$ 
(${\cal P}_{ab}(k)$) are the Green's functions for the scalar 
(pseudoscalar) mesons. 
${\cal S}_{08}(k)$ and ${\cal P}_{08}(k)$, however, are 
nonzero on account 
of the mixing between the singlet and the octet states. All 
other non--diagonal entries are identically zero. 
As such, 
it is necessary to rotate these Green's functions into the 
mass eigenbasis since only physical fluctuations can contribute to 
the masses: 
\begin{mathletters}
\bea 
U^{\rm S}_{i a} {\cal S}_{ab}(k) U^{\rm S}_{j b}
        &=&  \tilde{{\cal S}}_{i}(k) \delta_{i j} \\
U^{\rm P}_{i a} {\cal P}_{ab}(k) U^{\rm P}_{j b}
        &=&  \tilde{{\cal P}}_{i}(k) \delta_{i j} \,\, ,
\eea\label{eqn:mixingprops}
\end{mathletters}
where $U^{\pi}_{i a} = \delta_{i a}$ for $i, a \neq 0,8$
and where $U^{\pi}_{i a}$ is given by an $O(2)$ rotation 
by $\theta_{{\rm P}}$ in the 0--8 block. 
The definition for $U^{\sigma}_{i a}$ is similarly given with 
$\theta_{{\rm P}} \rightarrow \theta_{{\rm S}}$.
The thermal integral arising from tadpole diagrams is 
\bea
\int_k \tilde{{\cal S}}_{i}(k) = 
    \int \frac{d^3{\bf k}}{(2 \pi)^3}  
    \frac{1}{\epsilon_{\bf k}[(\tilde{M}^{2}_{S})_{i}]}
    \left(\exp\left\{\frac{\epsilon_{\bf k}[(\tilde{M}^{2}_{S})_{i}
        ]}{T}\right\}-1 \right)^{-1}\nonumber
\eea
and similarly for the pseudoscalar tadpole integrals, 
$\int_{k}  {\cal P}_{cd}(k)$.  Here, 
$\epsilon_{\bf k}[(\tilde{M}^{2}_{S})_{i}] = 
({\bf k}^{2} + 
(\tilde{M}^{2}_{S})_{i})^{1/2}$ is the relativistic energy of the $i$th 
scalar quasiparticle with momentum ${\bf k}$.
I have neglected the vacuum contribution arising from 
the loop integrals.  Implementing a systematic renormalization 
scheme is difficult but possible in this approximation  
(see \cite{JDRenorm}).  The results, however, are not 
significantly altered.

Since in the Hartree approximation, the gap 
equations do not have an explicit momentum dependence, 
we can assume that 
$({\cal S}_{ab}(k))^{-1} = -k^2 + M_{S}^2$
where $M_{S}$ depends on temperature but not momentum, 
and similarly for $({\cal P}_{ab}(k))^{-1}$. 
Equations (\ref{eqn:cond}) and (\ref{eqn:mixingprops}) 
are then fixed point equations and can be 
numerically solved simultaneously as a function 
of temperature for $M_{\pi}$, $M_{K}$, $M_{\eta}$, $M_{\eta'}$, 
$M_{\sigma}$, $M_{\kappa}$, $M_{f_{0}}$, $M_{a_{0}}$, 
$\bar{\sigma}_{0}$, $\bar{\sigma}_{8}$, 
$\theta_{{\rm P}}$ and $\theta_{{\rm S}}$.  
The numerical solutions for a variety of 
parameters are given in Ref.\ \cite{lrsb}.

The condensate and mass gap equations are solved with fixed $m$, $c$, 
$\lambda_{1}$ and $\lambda_{2}$, while varying 
the background fields, $h_{0}$ and $h_{8}$.
The determination of the coupling constants is 
detailed in Ref.\ \cite{lrsb}.  
For $c \neq 0$, 
the four couplings in the Lagrangian are fitted to 
yield the physical tree-level masses of the 
pion, kaon, $\sigma$, $\eta$ and $\eta'$, while
for $c=0$, the remaining three couplings are determined from 
the physical tree-level masses of the pion, kaon, $\sigma$ and $\eta$. 
The background fields are proportional to the 
current quark masses: $m_{\rm up} = m_{\rm down} = a 
(h_{0} + h_{8}/\sqrt{2})$, $m_{\rm strange} = b
(h_{0} - \sqrt{2} h_{8})$.  For simplicity, I assume temperature 
independent proportionality constants, $a$ and $b$. 
Requiring that $m_{\pi} = 138$ MeV, $m_{K} = 496$ MeV, 
$m_{\rm up} = m_{\rm down} = 10$ MeV and 
$m_{ \rm strange} = 150$ MeV, gives
$a=4.64 \times 10^{-6} \, [{\rm MeV}]^{-2}$ and  
$b=2.27 \times 10^{-6} \, [{\rm MeV}]^{-2}$.

\begin{figure}
\hspace*{0cm}
\mbox{\epsfig{file=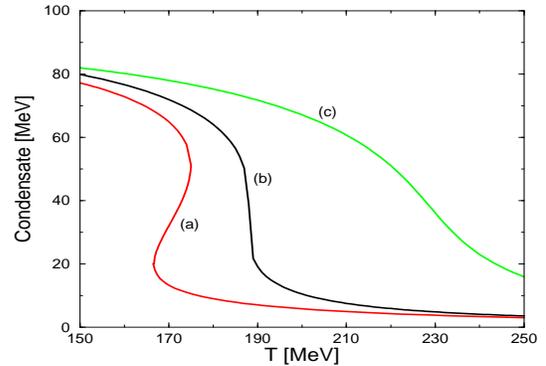,height=5cm,width=7cm,angle=0}}
\caption{The $\bar{\sigma}_{0}$ condensate for various 
values of the kaon mass with a pion mass of 100 MeV.  
The kaon mass is 80, 200 and 300 MeV for curves (a), (b) 
and (c), respectively.
Curve (a) is a first order phase transition, 
curve (b) is close to second order and curve 
(c) lies in the crossover region.}
\label{fig:condensates}
\end{figure}

To determine the order of the phase transition, I examined the 
continuity of the order parameters as a function of temperature. 
For a first order transition, the condensates are  
multivalued functions of temperature in the vicinity of the 
phase transition.  For a smooth crossover, the condensates 
are smooth singlevalued functions of temperature and always nonzero.
This behavior is demonstrated in Fig. \ref{fig:condensates}.  Only 
the nonstrange condensate is shown since both condensates 
exhibit qualitatively the same behavior.  

The numerical results are plotted in 
Fig.\ \ref{fig:phasediag2}.
For $c \neq 0$, these results agree with those of 
lattice groups \cite{Columbia,lattice2,lattice3}.
For $m_{\sigma} = 1000$ MeV, the authors of Ref.\ \cite{RobGavin} 
report
that the ratio of the critical current up-down quark mass to 
the physical up-down quark mass for 
$m_{\rm strange}/m_{\rm up,down} = 32$ to 
be $\sim 0.01$.  The corresponding value found in the present 
work is $\sim 0.20$.  The larger value found here is 
most likely due to the inclusion 
of thermal fluctuations from the scalar and pseudoscalar nonets.
For $m_{\rm up}=m_{\rm down}=0$, the transition is first order 
from zero strange quark mass to some critical strange quark 
mass.  
For $m_{\sigma} = 800$ MeV, this critical strange quark mass 
is about 16 MeV, while for $m_{\sigma} = 900$ MeV, it is 260 MeV.  

\begin{figure}
\hspace*{-0.5cm}
\mbox{\epsfig{file=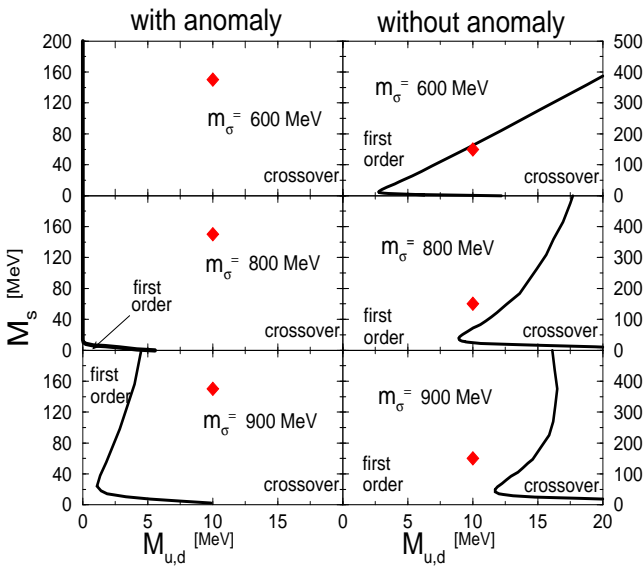,height=7.8cm,width=8.5cm,angle=0}}
\caption{The lines of second order 
phase transitions in the plane of the nonstrange and 
strange current quark masses for 
$m_{\sigma} = 600$ MeV, $m_{\sigma} = 800$ MeV and 
$m_{\sigma} = 900$ MeV.  The cases where the $U(1)_{A}$ symmetry 
is explicitly broken by the axial anomaly, 
$c \neq 0$, are shown on the left, and 
the cases where the $U(1)_{A}$ symmetry is exact, $c=0$, are 
shown on the right.
The physical 
mass point ($m_{\rm up,down} = 10$ MeV and 
$m_{\rm strange} = 150$ MeV) is indicated by the diamond. 
\label{fig:phasediag2}}
\end{figure}

For $c = 0$, however, the results are dramatically different. 
In particular, the line of second order transitions does not seem to
approach the strange quark mass axis.  For $m_{\sigma} = 600$ MeV, 
the physical point, $m_{\rm up} \cong m_{\rm down} \cong 10$ MeV and 
$m_{\rm strange} \cong 150$ MeV, is just outside the 
first order region.  For larger values of the $\sigma$ meson 
mass, the physical 
point is well within the first order region.  
The results also seem to indicate that for $c=0$ there is a first 
order phase transition for three flavors provided only that one
of the flavors is sufficiently heavier than the other two flavors. 
The departure of the second order phase transition line 
from the strange quark mass axis was also predicted using arguments from 
large-$N_c$ chiral perturbation theory in Ref.\ \cite{tytgat}.

At this point, it should be mentioned that 
the Hartree approximation sometimes
predicts a first order transition 
when the transition is actually second order. 
For example, renormalization group arguments predict a second order 
phase transition for the massless limit of the $O(4)$ linear 
sigma model \cite{PisarskiWilczek}, 
while the Hartree approximation predicts a 
first order transition (see, for instance, Ref.\ \cite{JDRenorm}).
This is not a problem in the low quark 
mass region since the transition is expected to be first order. 
The location of the second order line should 
not be significantly affected.

Additionally, 
the cubic and quartic couplings are fixed and temperature 
independent.  The running of the couplings with 
temperature should be at most logarithmic, while the 
integrals arising from the tadpole diagrams depend 
quadratically on the temperature.  So, it is reasonable 
that the running of the couplings does not 
qualitatively alter these results.
On the other hand, if the Coleman-Weinberg mechanism is 
strongly operative, some portion of the crossover region may 
actually be driven to first order \cite{Coleman}.

\begin{center} 
{\bf Acknowledgements} 
\end{center}
I especially want to 
thank Dirk H.\ Rischke, Robert D.\ Pisarski and J\"urgen 
Schaffner--Bielich for many valuable discussions.
I also want to thank the Nuclear Theory Group at Brookhaven 
National Laboratory for their generous support and 
hospitality while 
this work was completed. I am grateful to 
Mark Alford, Tom Blum, Eduardo Fraga, 
Robert Harlander, Alex Krasnitz,
and Ove Scavenius for useful discussions.  I am supported by the 
Director, Office of Energy
Research, Division of Nuclear Physics of the Office of 
High Energy and Nuclear Physics of the U.S.\ Department of 
Energy under Contract No.\ DE-AC02-98CH10886.

\end{narrowtext}

\end{document}